\documentclass[24pt]{article}
\usepackage{amssymb}
\usepackage{epsfig,caption,graphicx,eepic,epic}
\usepackage{pstricks}
\usepackage{amssymb,amsmath}
\usepackage{multido}
\usepackage{array}
\begin{document}
\def\be{\begin{equation}}
\def\ee{\end{equation}}
\def\ba{\begin{eqnarray}} 
\def\ea{\end{eqnarray}}
\def\nn{\nonumber}
\newcommand{\bbf}{\mathbf}   
\newcommand{\rrm}{\mathrm}
\title{\bf Quantum speed limit of a non-decoherent open system}

\author{Jean Richert$^{a}$
\footnote{E-mail address: j.mc.richert@gmail.com}\\ 
and\\
Tarek Khalil$^{b}$
\footnote{E-mail address: tkhalil@ul.edu.lb}\\ 
$^{a}$ Institut de Physique, Universit\'e de Strasbourg,\\
3, rue de l'Universit\'e, 67084 Strasbourg Cedex,\\      
France\\
$^{b}$ Department of Physics, Faculty of Sciences(V),\\
Lebanese University, Nabatieh,
Lebanon}


\date{\today}
\maketitle 

\begin{abstract}
We study the minimum time related to the quantum speed limit that characterizes the evolution of an open quantum system with the help of a simple model in the short and long time limits. We compare in particular the situation corresponding to a unique state and several states in the environment space. For short time intervals the results show a sensible difference in the behaviour of the system for different strengths of the interaction between the system and its environment.
 
\end{abstract}
\maketitle  

Keywords: open quantum systems, quantum speed limit, time scales\\

PACS numbers: 03.65.-w, 03.65.Yz, 05.30.Jp \\

\section{Introduction}


 The interval of time $\Delta t$ over which a system moves by an amount of energy $\Delta E$ has a long history which starts with the time-energy uncertainty relation derived by Heisenberg ~\cite{heis} and was further generalized ~\cite{rob1,rob2,sch}. Since then a considerable amount of work has developed from thereon, see ~\cite{dod}. It led to the derivation of different approximate estimations of the time related to  the so-called quantum speed limit $t_{QSL}$. The first important contributions concerning the optimisation of the time duration of quantum jumps have been realized by Mandelstam and Tamm ~\cite{mt} and followed by several other developments during the $90s$ ~\cite{ana,vaid,pfe,uff,ml}. The results concerned essentially closed systems.\\
 
During the following decade up to present days the subject developed further and extended to the case of open systems, see in particular ~\cite{dl1,dl2}. These developments were strongly motivated by the interest in the realization of fast electronic devices in which a given component interacts with its environment.\\ 

A large series of very recent work was devoted to the study of different properties which govern the behaviour of $t_{QSL}$ in such systems such as the structure of the system space ~\cite{pog},the environment space ~\cite{deh}, the importance of the interaction coupling with the environment and its optimization  ~\cite{mar,zha,khal1}, the non-Markovian memory effects ~\cite{xu,liu}, the initial state of the system ~\cite{jin}, the role of a thermal environment ~\cite{wu}, the influence of relativistic effects ~\cite{vil}.\\

In the present work we aim to consider the behaviour of the time related to the lower bound of $t_{QSL}$ and related quantities. We choose to perform a numerical analysis on a coherent open system described by a spin interacting with a bosonic environment. We first recall the expression of the time inequality in section 2 and 
develop the model and the analytic expression of the inequality in the framework of the model in section 3. Section 4 is devoted to the results. In section 5 we discuss and comment the results. Technical details are shown in the Appendices.
 
\section{The time-energy relation}
 
We introduce here the generalized energy-time relation to open quantum systems following the developments of Deffner and Lutz ~\cite{dl1,dl2}. 

Denote by $\hat \rho_{S}^{(0)}(t)$ the density operator of an isolated system $S$ and by
$\hat \rho_{S}(t)=Tr_{E}[\hat \rho(t)]$ the density operator of an open quantum system obtained as the trace over the states of an environment $E$ coupled to $S$, where $\hat \rho(t)$ is the density operator of the closed system 
$S \oplus E$. The system $S$ is characterized by its fidelity 

\ba
F(\hat \rho_{S}^{(0)}(0),\hat \rho_{S}(t))=Tr_{S}[(\hat \rho_{S}^{(0)}(0))^{1/2} \hat \rho_{S}(t) 
(\hat \rho_{S}^{(0)}(0))^{1/2})^{1/2}]^{2}
\label{eq1}
\ea
which measures at time $t$ the deviation of the open system from its initial state at the initial time at which the system does not interact with an environment. 

The deviation can be expressed in terms of an angle ~\cite{bur,joz} which reads

\ba
B(\hat \rho_{S}^{(0)}(0),\hat \rho_{S}(t))= \arccos( F(\hat \rho_{S}^{(0)}(0),\hat \rho_{S}(t))
\label{eq2}
\ea
which is the generalization in the case of mixed states of the angle characterizing the overlap of two states 
in Hilbert space.

In the general case the time evolution of the system $S$ is given by the master equation

\ba
d \hat \rho_{S}(t)/dt=D_{t}(\hat \rho_{S}(t))
\label{eq3}
\ea
where $D_{t}$ is a positive operator.

The average energy over a unit of time exchanged between $S$ and $E$ can be estimated by means of the quantity

\ba
<\Delta E(t)>=\frac{1}{t}\int_{0}^{t} dt' e(t')       
\label{eq4}
\ea
where $e(t)$ is taken as the smallest of the operator, trace or Hilbert-Schmidt norm of $D_{t}(\hat \rho_{S}(t))$: 

\ba
||D||_{op}(t)=\max [\lambda_{i}(t)] 
\notag\\
||D||_{tr}(t)=\sum_{i}[\lambda_{i}(t)] 
\notag\\
||D||_{hs}(t)=[\sum_{i}\lambda_{i}(t))^{2}]^{1/2}
\label{eq5}
\ea

The $[\lambda_{i}(t)]$'s are the time-dependent eigenvalues of $D_{t}(\hat \rho_{S}(t))$ which in practice is the time derivative of the known density operator $\hat \rho_{S}(t)$, Eq.(3).

The minimum time corresponding to the quantum speed limit obeys the inequality  given in  ref.~\cite{dl2}:

\ba
t \geq t_{LB}(=t_{QSL})
\label{eq6}
\ea
where the lower bound time is given by

\ba
t_{LB}=\hbar [\max(1/||D||_{op}(t),1/||D||_{tr}(t),1/||D||_{hs}(t))]
|\cos(B(\rho_{S}^{(0)}(0),\rho_{S}(t)))-1|
\label{eq7}
\ea
($\hbar=1$ in what follows).\\

This time corresponds to a lower limit of the evolution time of the system. It is the smaller the larger the denominator
corresponding to the energy exchange between $S$ and $E$ and the smaller the numerator, i.e. the larger the overlap beween the initial state and the state of the system at time $t$.

\section{Application of the inequality}

\subsection{Model}

The system $S$ is a spin which rotates around its $Oz$-axis and couples to a system of bosons.

The Hamiltonian $\hat H$ reads

\ba
\hat H= \hat H_{S}+\hat H_{E}+\hat H_{SE}
\label{eq8}
\ea

with 

\begin{center}
\ba
\hat H_{S}=\omega \hat J_{z} 
\notag \\
\hat H_{E}=\beta b^{+}b
\notag \\
\hat H_{SE}=\eta(b^{+}+b) \hat J^{2}
\label{eq9}
\ea 
\end{center}  
$(b^{+},b)$ are boson operators, $\omega$ is the rotation frequency of the system, $\beta$ the quantum of energy of the  bosonic oscillators and $\eta$ the strength parameter of the coupling interaction between $S$ and $E$.

In the present case $H_{SE}$ commutes with $H_{S}$ since $\hat J_{z}$ and $\hat J^{2}$ commute in the common basis of states $[|j m\rangle]$ which are the eigenstates of both operators. It has been shown ~\cite{lid} that this class of systems does not lead to decoherence. In the subspace of $\hat J_{z}$ limited to a unique spin state in $j$ the projection of $\hat H_{SE}$ on $S$ space is diagonal in $j_{1}$ and reads
  
\ba
\langle j_{1}|\hat H_{SE}|j_{2} \rangle=\eta j_{1}(j_{1}+1)(b^{+}+b)\delta_{j_{1},j_{2}} 
\label{eq10}
\ea 
As a consequence the density operator $\hat \rho_{S}(t)$ at time $t$ is obtained by taking the trace over the environment states of the total Hamiltonian $\hat \rho(t)$ leading to 

\ba
\hat \rho_{S}(t)=Tr_{E}\hat \rho(t) 
\label{eq11}
\ea 
whose matrix elements read  

\ba
\rho^{j_{1} m_{1}, j_{2} m_{2}}_{S}(t)=\rho^{j_{1} m_{1}, j_{2} m_{2}}_{0}(t)\Omega_{E}(j_{1},j_{2,}t)
\delta(j_{1},j_{2})
\label{eq12}
\ea  
with  
  
\ba
\rho^{(0) j_{1} m_{1}, j_{2} m_{2}}_{0}(t)=\frac{e^{[-i\omega(m_{1}-m_{2})]t}}{(\hat j_{1}\hat j_{2})^{1/2}}
\label{eq13} 
\ea  
where $\hat j_{i}=2j_{i}+1$, $m_{1},m_{2}$ are the spin projections on the quantization axis $Oz$. The bosonic environment contribution $\Omega_{E}$ can be put in the following form 
  
\ba
\Omega_{E}(j_{1},j_{2,};t)=1/N(t)\sum_{n=0}^{n_{max}}\frac{1}{n!}\sum_{n',n^{"}}\frac{E_{n,n'}(j_{1},t)
E^{*}_{n^{"},n}(j_{2},t)}{[(n'!)(n''!)]^{1/2}}
\label{eq14} 
\ea  
where $N(t)$ is a normalization factor such that the trace of the density operator in the total space $S \oplus E$ is constant in time, $Tr \hat \rho(t)=1$, $E_{n,n'}(t), E^{*}_{n'',n}(t)$ are scalar quantities related to the coupling between $S$ and $E$ and $n_{max}$ is the upper limit of the bosonic quantas present in the environment. An exact analytical expression of these quantities can be obtained by using an infinite Zassenhaus series ~\cite{za} whose development is given in Appendix A. 
The explicit expressions of the polynomials $E_{n_{1},n_{2}}(t)$ are developed in Appendix B.

By simple inspection of these expressions it can be seen that the non-diagonal element of $\rho^{j_{1} m_{1}, j_{2} m_{2}}_{S}(t)$ may cross zero with $t$, oscillate and never reach and stay at zero whatever the length of the time interval which goes to infinity. This is the signature of the fact that no decoherence of the system will be observable in this case.  

\subsection{The time-energy relation in the framework of the model}

In the application of the model the spin of the system is supposed to stay in a unique $j$ state, 
$j_{1}=j_{2}=j= 1/2$. The energy-time expression leads to the lower bound 

\ba
t \geq \frac {\hbar |\cos(B (\rho_{S}^{(0)}(0),\rho_{S}(t))-1|}{\Delta_{k}(t)}
\label{15}
\ea
which can be worked out analytically. The index $k$ corresponds to different choices of the integrant $\Delta(t)= <\Delta E(t)>$ 
which is chosen as the the smallest one given by  Eqs.(17 - 19). The fidelity takes the form

\ba
F^{j m_{1},j m_{2}}(\hat \rho_{S}^{(0)}(0),\hat \rho_{S}(t))=\frac{2 \Omega_{E}(j,j;t)(1+\cos(\omega t))} 
{\hat j^{2}}           
\label{16}
\ea
which is a real quantity and consequently the expression of $B(\rho_{S}^{(0)}(0),\rho_{S}(t))$ is obtained by means of Eq.(2). Finally the energy denominator $max(\Delta_{k}(t))$ where $[k=(op),(tr),(hs)]$ is given by 

\ba
\Delta_{op}(t)=1/t\int_{0}^{t} max[|\lambda_{i}(t')|] dt'
\label{17}
\ea
 
where $\lambda_{i}(t), i=1,2$ are the eigenvalues of the operator $d \hat \rho_{S}(t)/dt$ and similarly

\ba
\Delta_{tr}(t)=1/t\int_{0}^{t} [|\lambda_{1}(t')+|\lambda_{2}|(t')]dt'
\label{18}
\ea
and

\ba
\Delta_{hs}(t)=1/t\int_{0}^{t} [|\lambda_{1}(t')|^{2}+|\lambda_{2}(t')|^{2}]^{1/2} dt'
\label{19}
\ea

Details concerning the derivation of the fidelity are given in Appendix C.

\section{Numerical applications}

We use the present model in order to analyze the fidelity and the behaviour of the quantum speed limit under different physical conditions. We fix the energy parameters $\omega=1$ and $\beta=1$ and consider different strengths of the interaction $\eta_{w}=0.1$ (weak coupling=W), $\eta_{i}=1.0$ (intermediate coupling=I) and $\eta_{s}=5.0$ (strong coupling=S). We choose different extensions of the bosonic environment, in practice $n_{max}=0,5,10$. The system space is restricted to $j_{1},j_{2}=j=1/2$. At time $t=0$ the closed system 
is in a pure state, the state of lowest energy is  
\ba
|\psi_{S}(0) \rangle= 1/\sqrt2(|1/2 ; 1/2 \rangle + |1/2  ;-1/2 \rangle)
\label{20}
\ea
The bosonic contribution at $t=0$ is fixed by $\Omega_{E}(j,j;t=0)=1$.\\

We consider first the fidelity which is the essential ingredient in the r.h.s. of Eq.(15).

\subsection{Evolution of the fidelity: Bures angle}

a) Weak interaction: the fidelity shows an oscillating behaviour which indicates that the time dependent state vector $\psi_{S}(t)$ rotates with time. The evolution of $F(t)$ for different times $t$ is the following:


For $\eta_{w}$, $n_{max}=0$: $F(1)=0.39$, $F(1.5)=0.19$  and $F(t) \simeq 0$ for $t=15,20$. It oscillates for larger time intervals up to $t=200$ with sizable amplitudes comparable to those of small time intervals. 

The behaviour is qualitatively similar for $n_{max}=5$ and $n_{max}=10$. One observes a saturation effect  from $n_{max}=5$ to $n_{max}=10$, the values of the fidelity are very similar to each other.\\






b) Intermediate interaction: the results are similar to those observed in the weak case. The amplitude oscillations start with a similar decrease but diminish more quickly than in the former case.

For $\eta_{i}$, $n_{max}=0$: $F(1)=0.29$, $F(1.5)=10^{-4}$ and for $t=10,20$
$F(10)=2.2\times10^{-2}$, $F(20) \simeq 0$. 

For $n_{max}=5$: $F(1)=0.39$, $F(10)=2.0\times10^{-2}$. $F(20)=1.6\times10^{-1}$. The saturation effect from $n_{max}=5$ to $n_{max}=10$  is again seen here. The pseudo-period of oscillation is however somewhat shorter than in the weak case for $n_{max}=5$ and $n_{max}=10$. Oscillations persist over larger time intervals.\\





c) Strong interaction: the fidelity is large at the very beginning and gets quickly very small for early times. The oscillation period is much shorter than in the former cases. It goes on oscillating but with small amplitudes for larger time intervals.

For $\eta_{s}$, $n_{max}=0$: $F(0.15)=0.42$, $F(1)=2.0\times10^{-5}$, $F(2)=2.0\times10^{-6}$. 
It is similar for $n_{max}=5$: $F(0.15)=0.61$, $F(1)=2.0\times10^{-3}$, $F(2)=4.0\times10^{-4}$ and for
$n_{max}=10$: $F(0.15)=0.61$, $F(1)=2.0\times10^{-3}$, $F(2)=4.1\times10^{-4}$. The fidelity goes on oscillating over  
larger period of times. The saturation effect is also present here.\\






In summary the fidelity is an oscillating function of the length of the time interval for any strength of the interaction. It is large for small times with an overall decrease in amplitude with increasing time. Its pseudo-period decreases with increasing interaction strength between the system and its environment. One observes a saturation effect with an increasing number of states in the environment.

\subsection{The time dependent average energy of the system}

The average energy over a unit time interval $<\Delta E(t)>$ which enters the expression of the lower bound $t_{LB}$ of $t_{QSL}$ is defined in Eq.(4). The essential trend of its temporal evolution can be characterized  as shown below.\\

a)  Weak interaction: the average energy oscillates with a similar amplitude of the order of unity for small time intervals $t \in [0,20]$ and $n_{max}=0,5,10$. For large times ($t \in [50,200]$) $<\Delta E(t)>$ gets 
smaller and the oscillation amplitudes get reduced.\\

b) Intermediate interaction and strong interaction: the behaviour is qualitatively the same as in the weak case with smaller amplitudes and reduced oscillations over the time interval $t \in [0,200]$.

\subsection{The quantum speed bound}

The lower bound $t_{LB}$ defined in Eq.(7) is the physical quantity of interest in the present study. It depends on the behaviour of the fidelity (numerator) and the average energy per time unit present in the system (denominator). The analysis is done with the same set of parameters as those used for the fidelity.\\ 

a) Weak interaction: The lower bound $t_{LB}$ is quite smaller than the evolution time interval, 
$t_{LB} \leq t$, whatever the extension of the bosonic space $n_{max}=0,5,10$. It is very similar in magnitude in all cases. It is rather stationnary for $t \leq 5$ and then increases slowly and quasi linearly, see Fig.1. For larger times ($t \in [100,200]$)  $t_{LB}$ remains of the same order of magnitude and oscillates.\\
 
\begin{figure}
\epsfig{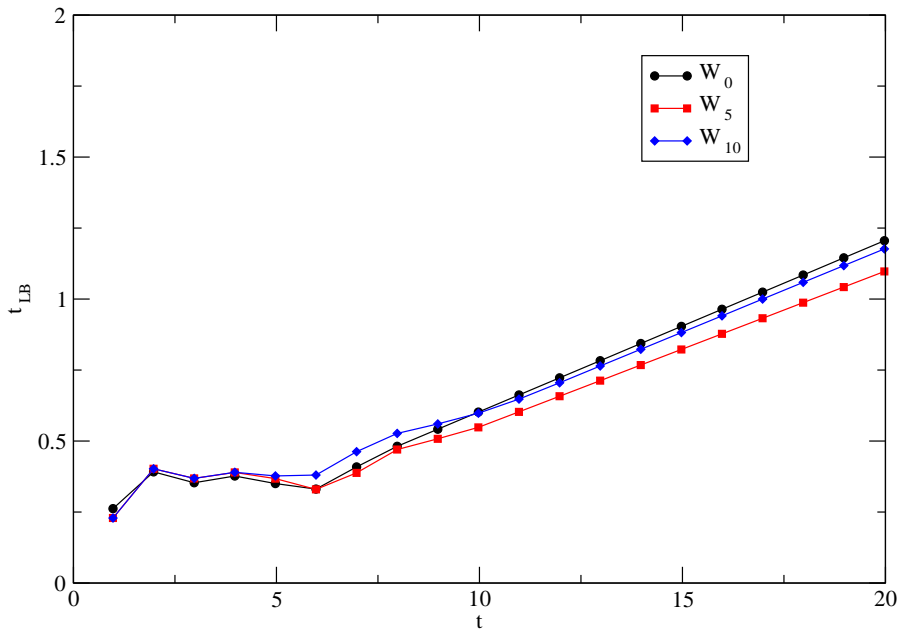}
\caption{The lower bound $t_{LB}$: weak interaction, $n_{max}=0,5,10$. See the text.}
\label{fig1}
\end{figure}

b) Intermediate interaction: 

For $n_{max}=0$ $t_{LB}$ is quite smaller than the time interval $t$ except for small $t \leq 5$ where it increases strongly with a steep slope, see Fig.2.\\

For $n_{max}=5,10$ one observes a saturation effect for which the values of  $t_{LB}$ come very close to each other. Otherwise the behaviour is the same as in the case $n_{max}=0$, except for the fact that the magnitude of the oscillations are smaller, see Fig.2. This remains so for larger time intervals. The magnitudes of $t_{LB}$ do not increase with time for $t \geq 5$ as it is the case in the weak regime, they tend even to decrease compared to their behaviour in the former regime.\\

\begin{figure}
\epsfig{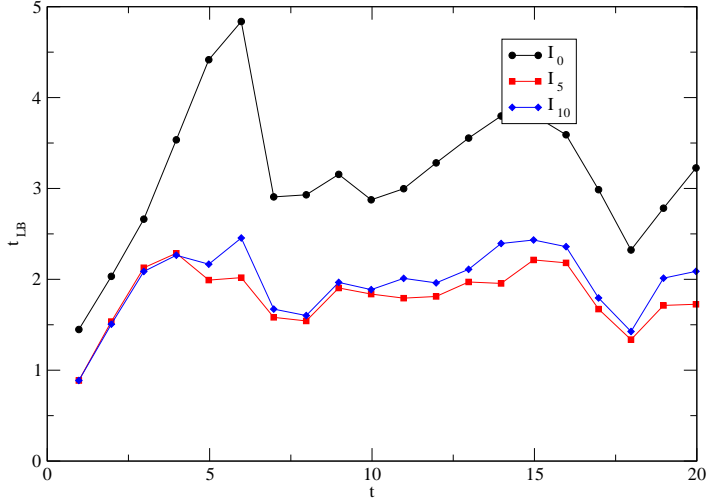}
\caption{The lower bound $t_{LB}$: intermediate interaction, $n_{max}=0,5,10$. See the text.}
\label{fig2}
\end{figure}

c) Strong interaction: 

Like in the intermediate strength case $t_{LB}$ is somewhat larger for $n_{max}=0$ than for $n_{max}=5,10$ but the oscillations in time follow rather closely in time and are more or less in phase with each other. The graph in Fig.3 shows sizable oscillations for $t \in [0,20]$. For larger times ($t \in [20,200]$) oscillations of $t_{LB}$ remain but they are smaller than those observed for small times. Like in the case of the fidelity, the difference between the case $n_{max}=5$ and $n_{max}=10$ is negligible, the curves cannot be distinguished from each other.\\

\begin{figure}
\epsfig{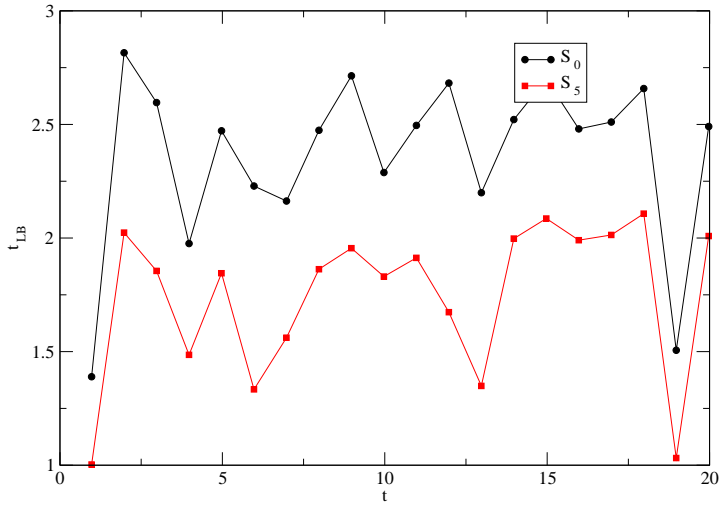}
\caption{The lower bound $t_{LB}$: strong interaction, $n_{max}=0,5$. See the text.}
\label{fig3}
\end{figure}

Summing up:

\begin{itemize} 

\item The lower bound of the time $t_{LB}$ is generally closer to the time interval $t$ for small time intervals than for long time intervals and shows oscillations which are essentially generated by the behaviour of the fidelity which appears in the numerator.


\item For all three interaction strengths there is no qualitative difference between the behaviour of $t_{LB}$ in the case where $n_{max}=0$ and 
$n_{max}=5,10$ except over short time intervals. The fact that $t_{LB}$ diminishes relative to increasing time intervals $t$ in the case of an intermediate and strong interaction is due to the fact that in both cases the fidelity $F(t)$ decreases and the energy denominator $<\Delta E(t)>$ varies slowly. This is not the case when the interaction is weak.\\   

\item The cases $n_{max}=5$ and $n_{max}=10$  are very similar, one observes a saturation effect with respect to an extension of the environment space in the size of $t_{LB}$.

\end{itemize}

\section{Final comments}

We used the time-energy inequality of Deffner and Lutz ~\cite{dl1,dl2,khal1} in order to examine the behaviour of the speed limit in a non-decoherent open quantum system ~\cite{khal2}.\\


The time which charaterizes the speed with which the open system evolves in time shows a lower bound $t_{LB}$ which is governed by the fidelity $F(t)$ and an accumulated average energy  per time unit $<\Delta E(t)>$. Both quantities oscillate in time. This is due to the nature of the density matrix of the spin $1/2$ system which shows  oscillatory matrix elements.\\

We considered an environment space which contains a set $n$ of non-interacting bosons with different numbers of excitation quantas and studied the evolution of the system for different values of the interaction strength between the system and the environment.\\ 

We first analyzed the behaviour of the fidelity $F(t)$ which leads to the Bures angle ~\cite{bur} and showed how the state vector of the open system evolves in time. It oscillates in the interval $[0,1]$ for all values of the interaction with different amplitudes and periods depending on the strength chosen to be weak, intermediate and strong. The oscillatory character is induced by the structure of the density operator of the system.
The oscillating average energy per time unit $<\Delta E(t)>$ varies due to the energy exchange between the system and its environment.\\

The physically observable of interest is the lower time limit $t_{LB}$. Different strengths lead to different lower time bounds $t_{LB}$.
In the case of a weak amplitude $t_{LB}$ is close to the time interval $t$ for small $t$.\\

For an intermediate and strong strengths of the interaction $t_{LB}$ comes closer to the time interval $t$ for small times. It gets smaller if the bosonic space contains several excited states. This is essentially due to the decrease of the numerator of the expression of $t_{LB}$ governed by the fidelity, to some extent to the variation of the energy $<\Delta E(t)>$. The size of $t_{LB}$ is the smallest when the interaction is weak.\\

As it can be seen on the figures, one observes a rather systematic "saturation" effect in the sense that  $t_{LB}$  starts to be insensitive to an enlargement of the environment space, i.e. it remains the same when one increases the number of excitation quanta from $5$ to $10$ excitations. This is related to the fact that higher contributions to the quantity $\Omega_{E}(t)$ (Eq.14) which enters the expression of the fidelity (Eq.16) get small when the number of excitation quanta $n$ increases.\\

It has been shown in a preceding work that the evolution of an open quantum system obeys the divisibility property which characterises a Markovian behaviour if the environment space reduces to a unique state ~\cite{khal3}. It would be of interest to analyze the behaviour of $t_{LB}$ in this specific case. 

\section{Appendix A: the Zassenhaus development} 
 
If $X=-i(t-t_{0})(\hat H_{S}+\hat H_{E})$ and $Y=-i(t-t_{0})\hat H_{SE}$

\ba
e^{X+Y}=e^{X}\otimes e^{Y}\otimes e^{-c_{2}(X,Y)/2!}\otimes e^{-c_{3}(X,Y)/3!}\otimes e^{-c_{4}(X,Y)/4!}...
\label{eq31}
\ea

where

\begin{center}
$c_{2}(X,Y)=[X,Y]$\\ 
$c_{3}(X,Y)=2[[X,Y],Y]+[[X,Y],X]$\\ 
$c_{4}(X,Y)=c_{3}(X,Y)+3[[[X,Y],Y],Y]+[[[X,Y],X],Y]+[[X,Y],[X,Y]$, etc.\\
\end{center} 
 
The series has an infinite number of term which can be generated iteratively in a straightforward way ~\cite{ca}. If $[X,Y]=0$ the truncation at the  third term  leads to the factorisation of the $X$ and the $Y$ contribution. If $[X,Y]=c$ where $c$ is a c-number the expression corresponds to the well-known Baker-Campbell-Hausdorff formula.  

  
Here the expressions of the evolution operator $\exp(-iHt)$ can be rigorously worked out analytically and the series can be formally summed up to infinity for any time by means of analytic continuation arguments.

\section{Appendix B: The bosonic content of the density operator}

The expressions of the bosonic contributions to the density matrix $\rho^{j_{1} m_{1}, j_{2} m_{2}}_{s}(t)$ are given by 
 
\ba
E_{n,n'}(j_{1},t)=e^{-i\beta t}\sum_{n\geq n_{2},n_{3}\geq n_{2}}\sum_{n_{3}\geq n_{4},n'\geq n_{4}}(-i)^{n+n_{3}}
(-1)^{n'+n_{2}-n_{4}}
\notag\\
\frac{n!n'!(n_{3}!)^{2}[\alpha_{1}(t)^{n+n_{3}-2n_{2}}][\zeta_{1}(t)^{n_{3}+n'-2n_{4}}]}{(n-n_{2})!(n_{3}-n_{4})!
(n_{3}-n_{2})!(n'-n_{4})!}e^{\Psi_{1}(t)}
\label{eq23}      
\ea
and

\ba
E^{*}_{n^{"},n}(j_{2},t)=e^{i\beta t}\sum_{n^{"}\geq n_{2},n_{3}\geq n_{2}}\sum_{n_{3}\geq n_{4},n\geq n_{4}}i^{n^{"}+n_{3}}
(-1)^{n+n_{2}-n_{4}}
\notag\\
\frac{n^{"}!n!(n_{3}!)^{2}[\alpha_{2}(t)^{n^{"}+n_{3}-2n_{2}}][\zeta_{2}(t)^{n+n_{3}-2n_{4}}]}{(n^{"}-n_{2})!(n_{3}-n_{2})!(n_{3}-n_{4})!(n-n_{4})!}e^{\Psi_{2}(t)}
\label{eq24}      
\ea

The different quantities which enter $E_{n,n'}(t)$ are 

\ba
\alpha_{1}(t)=\frac{\gamma(j_{1})\sin\beta t}{\beta}
\label{eq25}      
\ea

\ba
\zeta_{1}(t)=\frac{\beta[1-\cos\gamma(j_{1})t]}{\gamma(j_{1})}
\label{eq18}      
\ea

\ba
\gamma(j_{1})=\eta j_{1}(j_{1}+1)
\label{eq26}      
\ea

\ba
\Psi_{1}(t)=-\frac{1}{2}[\frac{\gamma^{2}(j_{1})\sin^{2}(\beta t)}{\beta^{2}}+\frac{\beta^{2}(1-\cos\gamma(j_{1})t)^{2}}{\gamma^{2}(j_{1})}]                           
\label{eq27}      
\ea

and for $E^{*}_{n'',n}(t)$: 
\ba
\alpha_{2}(t)=\frac{\gamma(j_{2})\sin\beta t}{\beta}
\label{eq28}      
\ea

\ba
\zeta_{2}(t)=\frac{\beta[1-\cos\gamma(j_{2})t]}{\gamma(j_{2})}
\label{eq29}      
\ea

\ba
\gamma(j_{2})=\eta j_{2}(j_{2}+1)
\label{eq30}    
\ea

\ba
\Psi_{2}(t)=-\frac{1}{2}[\frac{\gamma^{2}(j_{2})\sin^{2}(\beta t)}{\beta^{2}}+\frac{\beta^{2}(1-\cos\gamma(j_{2})t)^{2}}{\gamma^{2}(j_{2})}]                           
\label{eq31}      
\ea

\section{Appendix C: Calculation of the fidelity in the framework of the model}

Define the operator

\ba
M(t)=[((\hat \rho_{S}^{(0)}(0))^{1/2} (\hat \rho_{S}(t)) (\hat \rho_{S}^{(0)}(0))^{1/2})^{1/2}]^{2}
\ea 
Then the density operator in $S$ space reads:         

\ba
\hat \rho_{S}(t)=\frac{\hat \rho^{(0)}_{S}(t)*\Omega_{E}(j,j,t)}{(2j+1)}
\ea

For $j=1/2$ the elements of the $2*2$ matrix $M(t)$ can be worked out and read 

\ba
M^{(m_{1},m_{2})}(t)=Re(\rho_{S}^{(1)}(t)+\rho_{S}^{(2)}(t)
\ea

where 

\ba
\rho_{S}^{(1)}(t)=\rho_{S}^{(-1/2,-1/2)}(t)=1                
\notag\\
\rho_{S}^{(2)}(t)=\rho_{S}^{(-1/2,+1/2)}(t)=\exp(i\omega t)
\ea

It is easy to construct the matrix $(M^{(m_{1},m_{2})}(t))^{1/2}$ whose matrix elements read

\ba
M^{(m_{1},m_{2})}(t))^{1/2}=\frac{[\Omega_{E}(j,j,t)(1+\cos \omega t)]^{1/2}}{2^{1/2}(2j+1)}
\ea

Thereafter the fidelity  $F(\hat \rho_{S}^{(0)}(0),\hat \rho_{S}(t))$ can be  obtained and reads

\ba
F(\hat \rho_{S}^{(0)}(0),\hat \rho_{S}(t))=\frac{2 \Omega_{E}(j,j,t)(1+\cos \omega t)}{(2j+1)^{2}}
\ea

\end{document}